\documentclass[aps, 10pt, twocolumn,prl,superscriptaddress,showpacs, showkeys]{revtex4-1}
\usepackage[colorlinks,bookmarks=false,citecolor=blue,linkcolor=red,urlcolor=blue]{hyperref}
\usepackage{graphicx,epstopdf,color}
\usepackage{amsfonts}
\usepackage{amsmath,amssymb,mathrsfs}
\usepackage{bm}
\usepackage{pst-grad}

\def\bc{\begin{center}}
\def\ec{\end{center}}
\newcommand{\bs}[1]{\boldsymbol{#1}}
\newcommand{\ket}[1]{\left|#1\right\rangle}

\newcommand{\up}{\uparrow}
\newcommand{\dw}{\downarrow}

\newcommand{\bea}{\begin{eqnarray}}
\newcommand{\eea}{\end{eqnarray}}

\def\ea{\emph{et~al.}}

\newcommand{\be}{\begin{equation}}
\newcommand{\ee}{\end{equation}}

\newcommand{\bfss}{{\boldsymbol{S}}}

\newcommand{\bfx}{{\boldsymbol{x}}}
\newcommand{\bfy}{{\boldsymbol{y}}}

\def\bmx{\begin{pmatrix}}
\def\emx{\end{pmatrix}}
\begin{document}

%%%%%%%%%%%%%%%%%%%%%%%%%%%%%%%%%%%%%%%%%%%%
%\title{Correlated topological insulators in cold-atom gases with non-Abelian gauge fields}
%\title{Interaction effects in the Hofstadter-Hubbard model}%spinful time-reversal-invariant Hofstadter problem}
%\title{Interaction effects in the spinful time-reversal-invariant Hofstadter problem}
\title{Time-Reversal-Invariant Hofstadter-Hubbard Model with Ultracold Fermions}
%\title{Correlated phases in the cold-atom Hofstadter-Hubbard model}
%\title{Interaction effects in Topological insulators in cold-atom gases with non-Abelian gauge fields}
%%%%%%%%%%%%%%%%%%%%%%%%%%%%%%%%%%%%%%%%%%%%
%
\author{Daniel Cocks}
\altaffiliation{These authors contributed equally to this work.}
\affiliation{Institut f\"ur Theoretische Physik, Goethe-Universit\"at, 60438 Frankfurt/Main, Germany}
\author{Peter P. Orth}
\altaffiliation{These authors contributed equally to this work.}
\affiliation{Institute for Theory of Condensed Matter, Karlsruhe Institute of Technology (KIT), 76131 Karlsruhe, Germany }
%\affiliation{Institute for Theory of Condensed Matter, Karlsruhe Institute of Technology (KIT), 76131 Karlsruhe, Germany }
%
\author{Stephan Rachel}
\affiliation{Department of Physics, Yale University, New Haven, Connecticut 06520, USA}
\affiliation{Institute for Theoretical Physics, Dresden University of Technology, 01062 Dresden, Germany}
\author{Michael Buchhold}
\affiliation{Institut f\"ur Theoretische Physik, Goethe-Universit\"at, 60438 Frankfurt/Main, Germany}
\author{Karyn Le Hur}
\affiliation{Department of Physics, Yale University, New Haven, Connecticut 06520, USA}
\affiliation{Center for Theoretical Physics, Ecole Polytechnique, CNRS, 91128 Palaiseau Cedex, France}
\author{Walter Hofstetter}
\affiliation{Institut f\"ur Theoretische Physik, Goethe-Universit\"at, 60438 Frankfurt/Main, Germany}

\begin{abstract}
We consider the time-reversal-invariant Hofstadter-Hubbard model which can be realized in cold atom experiments. In these experiments, an additional staggered potential and an artificial Rashba--type spin-orbit coupling are available. Without interactions, the system exhibits various phases such as topological and normal insulator, metal as well as semi--metal phases with two or even more Dirac cones. Using a combination of real-space dynamical mean-field theory and analytical techniques, we discuss the effect of on-site interactions and determine the corresponding phase diagram. In particular, we investigate the semi--metal to antiferromagnetic insulator transition and the stability of different topological insulator phases in the presence of strong interactions. We compute spectral functions which allow us to study the edge states of the strongly correlated topological phases.
\end{abstract}

\maketitle

% stress the fact that synthetic gauge fields in cold atoms are fashionable
% Unique to optical lattices is the possibility to create new quantum systems: i.e. spin-dependent magnetic fields, spin-orbit coupling, large-N version of graphene

% New Organization:
% spinful Hofstadter problem + U (fundamental Hamiltonian, large-N version of graphene)
% Tunable magnetic interaction with additional gauge field terms available in experiments (DM) 
% spinful Hofstadter problem gives rise to topological insulator phases -> TI + U, in particular: stability of the phases, drive NI-TI transistion via U, fate of the edge states, detection and stability w.r.t. trapping potential (in suppl. material) 
\emph{Introduction}.---
Ultracold quantum gases trapped in optical lattice potentials provide insight into strongly correlated condensed matter systems. Examples are the Mott-insulator-superfluid transition, the dynamics of the Hubbard model after a quench of parameters, and the simulation of quantum magnetism~\cite{ColdAtomsCondMat}. Striking is the precise experimental control over almost all system parameters, including the particle-particle interaction strength. Simulating more traditional electronic condensed matter systems, however, is complicated by the fact that cold atoms are charge neutral, and their center of mass motion is thus not affected by external magnetic or electric fields (apart from trapping potentials). An experimental breakthrough was thus the engineering of so-called ``artificial'' gauge fields, which give rise to effective magnetic or electric fields for the neutral particles~\cite{ArtGauge}. Remarkably, they may even be generalized to simulate spin-orbit couplings or non-Abelian fields~\cite{SOC-OL}. 
%lin_spin-orbit-coupled_2011,PhysRevLett.107.255301,Spielman_PeierlsPhasearXiv_2012,StruckSengstock-GaugePotential-arXiv_2012}. 
The effective electromagnetic fields and couplings can be large, which allows, for example, realization of the quantum (spin) Hall effect in a completely new experimental context~\cite{kane-mele, ti-reviews, bernevig-PRL-2006, goldman-10prl255302, gerbier-10njp033007}. 

The underlying idea of realizing time-reversal-invariant two-dimensional (2D) topological phases with cold atoms is as simple as it is fundamental~\cite{bernevig-PRL-2006, goldman-10prl255302}. Consider the (integer) quantum Hall effect (QHE) on a 2D square lattice where an external magnetic field along the $z$ direction breaks time-reversal and translational symmetry. The single particle spectrum for arbitrary magnetic field strength -- having the shape of a butterfly -- was first computed by Douglas Hofstadter~\cite{hofstadter76prb2239}
%{hofstadter_energy_1976}, 
since then referred to as the {\it Hofstadter butterfly}. If the magnetic flux per plaquette is a rational number $\alpha=p/q$, in units of the Dirac flux quantum $\Phi_0 = h/e$, the system remains translationally invariant with an enlarged unit cell of $q$ lattice sites. The spectrum consists of $q$ energy bands and in all energy gaps one finds a finite Chern number $C$ and correspondingly $|C|$ chiral edge modes per edge. Interestingly, for even values of $q$ the system is a semi-metal at half-filling and exhibits $q$ Dirac cones.

To restore time-reversal symmetry we can imagine applying a magnetic field in the $z$ direction that only couples to the up-spins and a second field of the same strength but opposite direction that only couples to the down-spins. We thus end up with a spinful and time-reversal-invariant (TRI) version of the fundamental Hofstadter problem. Remarkably, such a scenario is feasible using cold atoms in artificial gauge fields~\cite{goldman-10prl255302,gerbier-10njp033007}. %where the opposing magnetic fields are created using artificial gauge fields. 
Thus, the semi-metallic Dirac dispersion for even $q$ becomes a generalization of graphene with a tunable number of Dirac cones. 
%constitutes a natural large-$N$ version of graphene. 
Energy gaps which were crossed by a single chiral edge mode in the QHE setup are now traversed by a helical Kramer's pair of edge states, corresponding to a topological insulator phase. Note that one can use the same Gedanken experiment to construct the Kane--Mele model~\cite{kane-mele}
%{PhysRevLett.95.146802,PhysRevLett.95.226801} 
from two time-reversed copies of Haldane's honeycomb model~\cite{haldane88prl2015}. 
%{PhysRevLett.61.2015}. 
The Kane--Mele model with additional Hubbard interaction has recently been intensively studied~\cite{KMH},
%~\cite{PhysRevB.82.075106,PhysRevLett.106.100403, zheng-ArXiv2011, PhysRevLett.107.166806, PhysRevLett.107.010401, WuRachelLeHur-QSHArXiv, PhysRevLett.108.046401,PhysRevB.85.045123, hohenadler-arXiv2011, Mardani-ArXiv2011}, 
in contrast to the Hofstadter~problem.

\begin{figure}[b]
\centering
\includegraphics[width=\linewidth]{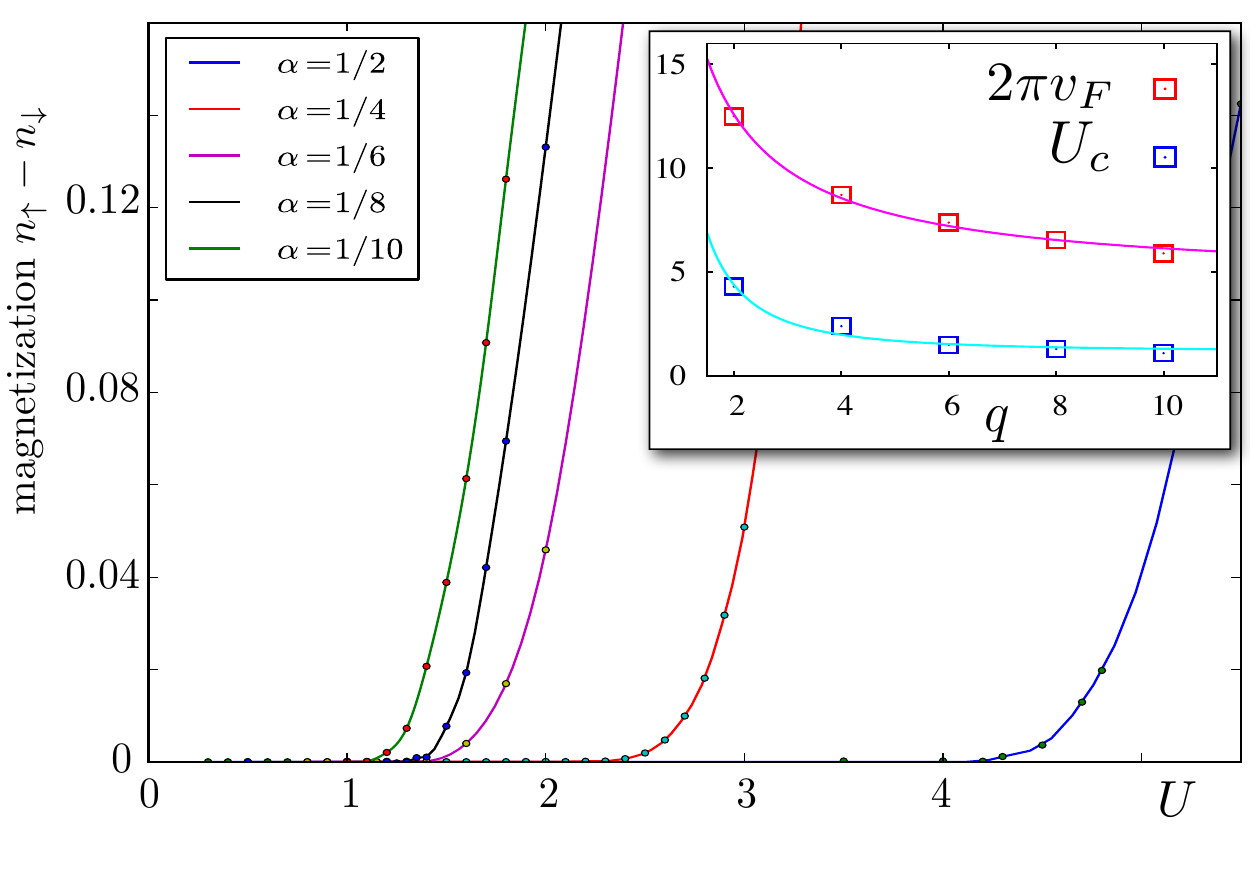}
\caption{(color online). Magnetization $m=n_{\up}-n_{\dw}$ in the N\'eel state is plotted versus interaction strength $U$. We show results for $\alpha=1/2$ (blue), $1/4$ (red), $1/6$ (magenta), $1/8$ (black), and $1/10$ (green) (curves from left to right). Inset: Fermi velocity $2\pi v_F$ (red symbols) for different $\alpha=1/q$ is shown versus $q$. $U_c$ (blue symbols) obtained within RDMFT versus $q$ is also shown. Magenta (upper) line is a fit of $v_F$ to $\propto 1/q$ and cyan (lower) line of $U_c$ to $\propto 1/q^2$. Note that odd $q$ denominators exhibit $U_c = 0^+$ due to a nested Fermi surface. }
\label{fig:Uc-alpha}
\end{figure}

\begin{figure}[b]
\centering
\includegraphics[width=\linewidth]{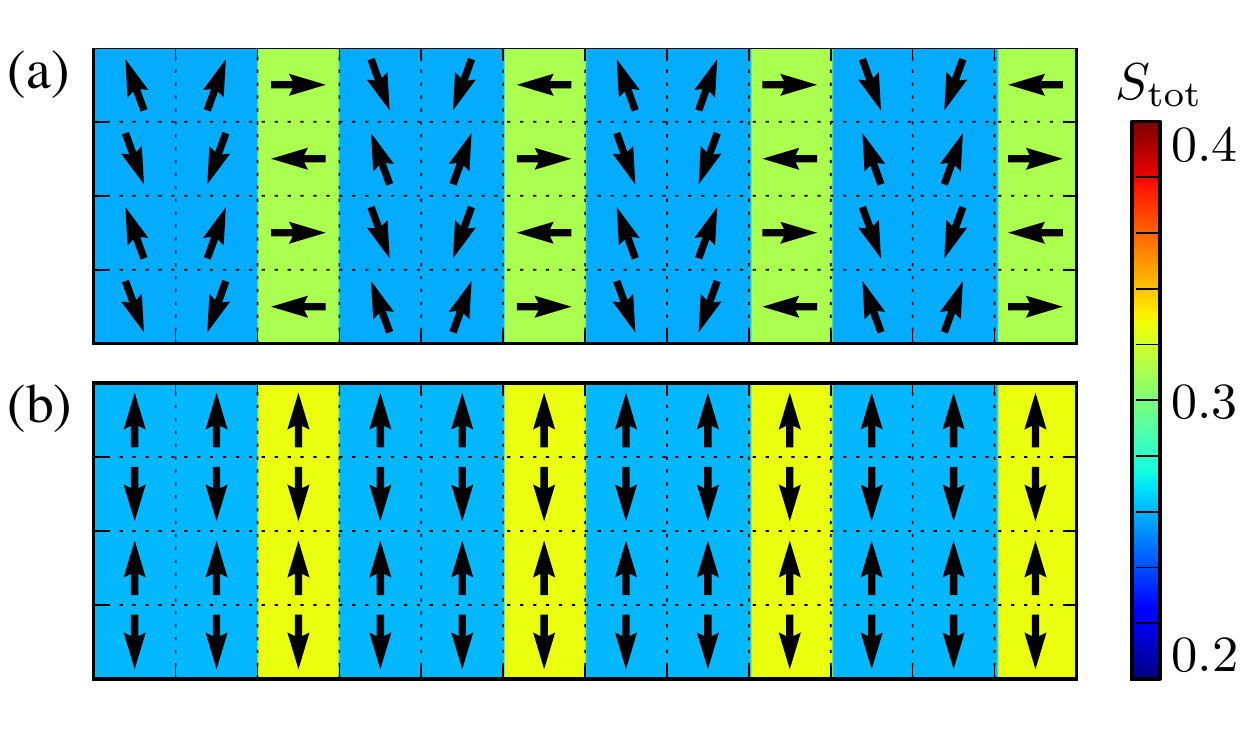}
\caption{(color online). Real space magnetization profile $\bs{m}(\bs{x})$ in $S^y$--$S^z$ plane for $\alpha=1/6$, $U=5$, $\lambda_x=0$, and $\gamma=0.125$ (a) and $\gamma=0.25$ (b), respectively.}
\label{fig:02}
\end{figure}

In this Letter we study the effect of interactions in the (TR-invariant) \emph{Hofstadter-Hubbard model} using real-space dynamical mean-field theory (RDMFT)~\cite{RDMFT_refs}. We explain our numerical results using analytical arguments. We consider interaction effects on both (semi-)metallic and gapped topological phases. Although $\mathbb{Z}_2$ topological insulators are known to be robust against disorder~\cite{moore-07prb121306, disorder},
% %~\cite{PhysRevB.75.121306, PhysRevB.78.195125, PhysRevLett.105.115501} 
rigorous and general results about the fate of topological insulators in the presence of Coulomb or Hubbard interactions are limited~\cite{ti+int}. Some three--dimensional materials of the iridate family are possible candidates for systems where strong spin--orbit coupling and Coulomb interactions compete~\cite{pesin-10np376,kargarian-11prb165112}.
% %~\cite{PesinBalents_TMI,PhysRevB.83.165112}. 
In two dimensions, however, topological insulator phases have so far only been found in HgTe/CdTe quantum wells~\cite{bernevig-06s1757,koenig-07s766}, where Coulomb interactions seem to be negligible.
% %~\cite{bernevig_quantum_2006,koenig_quantum_2007}. 
%In these quantum wells Coulomb interactions seem to be negligible. 

%In the semi-metal, interactions drive a transition to a magnetic insulator above a critical interaction strength, a situation similar to graphene. Inclusion of a Rashba-like spin-orbit coupling term available in experiments leads to tunable magnetic order. We explicitly show that the topological phases are robust with respect to interactions and lattice perturbations. We calculate the interacting phase diagram, and determine the fate of the topologically protected gapless edge modes, which are key to the experimental detection of the topological phase. 

\emph{Interacting TRI Hofstadter problem}.---
The TRI Hofstadter-Hubbard model is described by the Hamiltonian
\begin{align}
\label{eq:3}
  H =&  - \sum_{j}  \Bigl\{ t_x c^\dag_{j+\hat{\bfx}} c_{j} + t_y c^\dag_{j + \hat{\bfy}} \,e^{i 2 \pi \alpha x \sigma^z} c_j \nonumber \\ \qquad \qquad &+ \text{h.c.} \Bigr\} +\sum_j  U n_{j, \uparrow} n_{j, \downarrow} \,,
\end{align}
where $c_j^\dag=(c^\dag_{j\up}, c^\dag_{j\dw})$ at lattice site $j = (x,y)$, $\sigma^z$ is a Pauli matrix and $\hat{\bfx}=(1,0)$, $\hat{\bfy}=(0,1)$ are unit vectors. $t_{x}$ ($t_y$) is the hopping amplitude in $x$($y$--)~direction. We focus on isotropic hopping $t_x = t_y = t$ here, and express all energies in units of $t \equiv 1$. The value of $\alpha$ determines the strength of the (artificial) magnetic field for either spin species which penetrates a lattice plaquette in units of the Dirac flux quantum. The on-site interaction strength $U$ can be experimentally tuned by Feshbach resonances and by adjusting the lattice depth. For $U=0$ this model was studied in Ref.~\onlinecite{goldman-10prl255302} (for experimental details see Supplemental Material).  
\begin{figure*}[t]
\centering
\includegraphics[width=\linewidth]{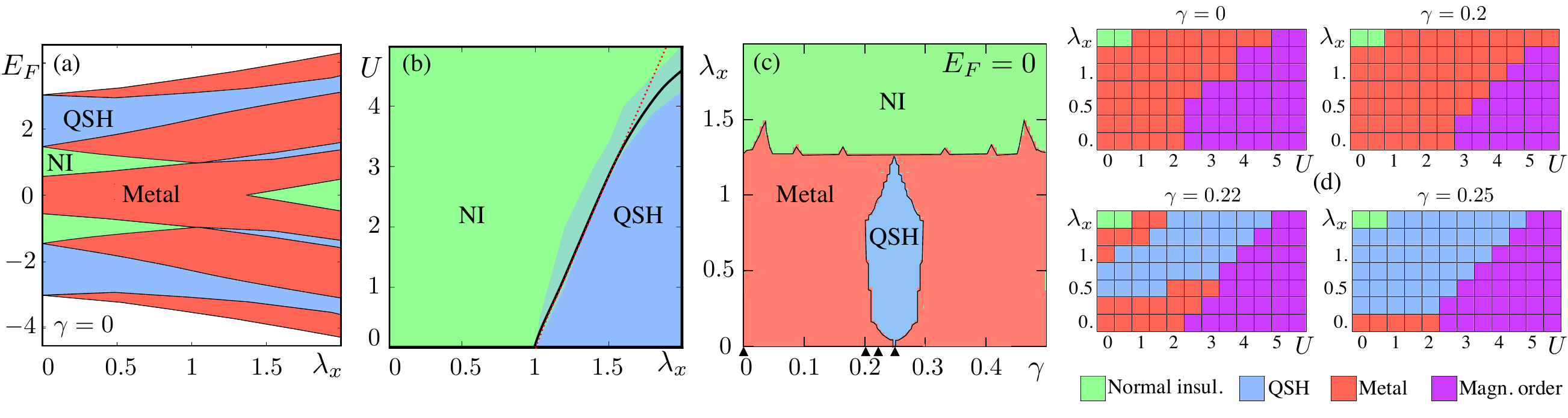} 
\caption{(color online). (a) $E_F$--$\lambda_x$--phase
diagram at $\gamma=U=0$. (b) $U$--$\lambda_x$--phase diagram at
$n_F = 2/3$ and $\gamma=0$. (c) $\lambda_x$--$\gamma$--phase diagram at
half filling $n_F=1$ and $U=0$. (d) $U$--$\lambda_x$--phase diagrams at $E_F=0$ for
various values of $\gamma$ [indicated by arrows in part (c)] and inverse temperature $\beta = 20$. We find (semi)metal (red), normal
insulator (NI) (green, in upper left), topological insulator (QSH) (blue), and
magnetically ordered phases (purple, in lower right).}
\label{fig:a16-U0}
\end{figure*}

%In the following we investigate the effect of a Hubbard on-site interaction $H_I=U\sum_i n_{i\up}n_{i\dw}$ which can be experimentally tuned using Feshbach resonances. 
We first consider the TRI Hofstadter-Hubbard problem for general $\alpha=p/q$ at half filling. % and $\gamma = \lambda_x = 0$. 
For $q$ odd the system is metallic with a nested Fermi surface, and antiferromagnetic N\'eel order occurs for infinitesimally small interaction $U=0^+$ as for the ordinary square lattice. For $q$ even the situation is very different because the system is a semimetal (SM) at half filling~\cite{Kohmoto-PRB-1989}. The noninteracting band structure exhibits $q$ Dirac cones (with a multiplicity of 2 due to spin) which are separated by momentum $2\pi/q$ in momentum space. The $\alpha=1/2$ case is thus very similar to graphene (but note that the coordination number is $z=4$ rather than $z=3$). For smaller $\alpha$ on the other hand the system embodies a generalization of graphene with a tunable number of valleys. 
%{\it natural} large-$N$ version of graphene, where $N$ refers to the number of valleys. 

We investigate the SM--insulator transition for various $\alpha=1/q$ ($q$ even) within RDMFT. In Fig.~\ref{fig:Uc-alpha}, the magnetization is shown as a function of interaction $U$. The insulating phase for $U>U_c$ is antiferromagnetically (AFM) ordered with a magnetization pointing in the $z$--direction and an ordering wave vector $\bs{Q}=(\pi,\pi)$. %Note that the N\'eel transition temperature is strictly zero in this 2D system. 
%a real--space dynamical mean--field theory (RDMFT)~\cite{RDMFT_refs} and 
We find that the critical value of $U_c$ to enter the insulating and magnetically ordered phase decreases for increasing $q$. This is expected from the increased scattering that can take place between the cones. At $U_c$ we also observe a simultaneous opening of the single particle gap. Within our approach we thus find no sign of an intermediate nonmagnetic gapped phase. 

To understand the behavior of $U_c(q)$ we make use of Herbut's argument~\cite{herbut-06prl146401}. 
%{PhysRevLett.97.146401} 
%and slightly modify it in order to explain the behavior of $U_c$ found numerically. 
Herbut considers graphene and studies the SM--insulator transition within a large-$N$ approach, and finds that $U_c$ depends on $2N$, the number of Dirac cones ($N$ refers to the spin degeneracy), and the Fermi velocity $v_F$ as $U_c \sim v_F/2 N$. As shown in detail in the Supplemental Material, we are able to match our results with Herbut's analysis by replacing the Fermi velocities and $2N=qN$. In fact, from the bandstructure at $U=0$ we find $v_F \propto 1/q$. Consequently, setting $N = 2$ for spin-1/2 particles, $U_c$ should exhibit a $1/q^2$ behavior which agrees very well with the RDMFT data; see inset of Fig.\,\ref{fig:Uc-alpha}. We further note that we find that $U_c (\alpha = 3/8) < U_c(\alpha = 1/8)$, which is in agreement with the analysis above since $v_F(\alpha=3/8) < v_F(\alpha = 1/8)$.

\emph{Specific cold atom setup}.---
We now consider two additional terms in the Hamiltonian that are available in the cold-atom setup~\cite{goldman-10prl255302, gerbier-10njp033007}: a staggering of the optical lattice potential along the $x$ direction 
\begin{equation}
  \label{eq:1}
  H_\lambda = \sum_{j} (-1)^x \lambda_x c^\dag_{j} c_j\,,
\end{equation}
and a Rashba-like spin-orbit coupling that breaks axial spin symmetry. It is introduced via replacing in Eq.~\eqref{eq:3}
\begin{equation}
  \label{eq:2}
  t_x \rightarrow t_x \exp( - i 2 \pi \gamma \sigma^x) \,.
\end{equation}
We first study the effect of finite $\lambda_x$ and $\gamma$ on the magnetic ordering. 

\emph{Tunable magnetism}.---
For $\gamma=0$, increasing $\lambda_x$ increases $U_c$ but does not change the type of magnetic order. Finite $\gamma$ does, however, change the type of magnetic order in general. To demonstrate this we consider fixed $U=5$ at $\alpha = 1/6$ and calculate the magnetization in real space for various values of $\gamma \in [0,0.25]$. The spin operators are defined in terms of the fermionic operators as usual as $\bfss_j = \frac12 c^\dag_j \boldsymbol{\sigma} c_j$ with $\boldsymbol{\sigma} = (\sigma^x, \sigma^y, \sigma^z)$. We show the magnetization pattern for $\gamma = 0.125$ and $\gamma = 0.25$ in Fig.~\ref{fig:02} obtained within RDMFT. We obtain similar results for other values of $\alpha$ and $\gamma$. For $\gamma = 0.125$, the magnetization lies in the $S^y$--$S^z$ plane, has a periodicity of six (two) lattice sites along $x$($y$) and reads explicitly $\bs{m}(\bs{x}) = S_{\text{tot}}(x) \cos \pi y \bigl( 0, -\cos(\frac{\pi x}{3} + \eta) , \sin (\frac{\pi x}{3} + \eta) \bigr)$ with $\eta = 0.39 \sin\frac{\pi x}{3} \cos \frac{\pi x}{3}$. The magnetization makes angles of $\{0^\circ, 70^\circ, 110^\circ, 180^\circ, 250^\circ, 290^\circ\}$ in the $S^y$-$S^z$ plane (spiral order). For $\gamma = 0.25$, the magnetic order is given by $\bs{m}(\bs{x}) = S_{\text{tot}}(x) ( 0, 0, \cos \pi y)$ (collinear order). Quantum fluctuations reduce the size of the magnetization $S_{\text{tot}} < 1/2$, which depends not only on the parameters $\alpha, \gamma$ and $U/t$, but is also spatially staggered for intermediate values of $U/t$ (see Fig.~\ref{fig:02}). The staggering decreases for larger values of $U/t$. More importantly, tuning the parameter $\gamma$ we pass from N\'eel to spiral to collinear order crossing two magnetic quantum phase transitions. %Finally, we note that the magnetization amplitude $S_{\text{tot}}$ is staggered for this intermediate value of $U$. The staggering decreases for larger values of $U$. 

We can qualitatively understand this type of magnetic order by rigorously deriving a quantum spin Hamiltonian for even stronger interactions when charge fluctuations freeze out at half filling (see Supplemental Material for details)
\begin{align}
  \label{spin-ham}
\mathcal{H} &= J_x \sum_j \biggl\{  S^x_j S^x_{j+\hat{\bfx}} + \cos (4 \pi \gamma) \Bigl[S^y_{j} S^y_{j+\hat{\bfx}} + S^z_{j} S^z_{j+\hat{\bfx}} \Bigr] \nonumber \\ &+ \sin( 4 \pi \gamma ) \Bigl[ S^z_j S^y_{j+\hat{\bfx}} - S^y_{j} S^z_{j+\hat{\bfx}}  \Bigr] \biggr\} \nonumber \\ & + J_y \sum_{j} \biggl\{ \cos \bigl( 4 \pi \alpha x \bigr) \Bigl[ S^x_{j } S^x_{j+\hat{\bfy}} + S^y_{j} S^y_{j+\hat{\bfy}} \Bigr] + S^z_{j} S^z_{j+\hat{\bf y}} \nonumber \\
&+ \sin( 4 \pi \alpha x) \Bigl[ S^y_j S^x_{j + \hat{\bfy}} - S^x_j S^y_{j + \hat{\bfy}}  \Bigr] \biggr\} 
\end{align}
where $J_i=4t_i^2/U$. The first part describes spin exchange in $x$ direction. For $\gamma= n/2$ with $n \in \mathbb{Z}$ we obtain a simple antiferromagnetic Heisenberg term. Other values of $\gamma$, however, break SU(2) symmetry and cause anisotropy of the $XXZ$ type with $S^x$ as anisotropy direction in spin space. For $\gamma \neq n/4$ there is an additional Dzyaloshinskii-Moriya (DM) interaction term in the $YZ$ plane, which is responsible for the spiral spin order in Fig.~\ref{fig:02}(a). Spin exchange in the $y$ direction is periodic with an extended unit cell in the $x$ direction depending on the flux $\alpha=p/q$: for odd $q$ the unit cell contains $q$ lattice sites, but for even $q$ it only contains $q/2$ lattice sites, reflecting second order perturbation theory. For instance, one finds for the $\pi$--flux lattice ($\alpha=1/2$) an ordinary Heisenberg exchange term. For other values of $\alpha$ the $XY$ term exhibits a modulation of its amplitude depending on $\alpha$, while the $Z$ term always favors AFM Ising order. The rich magnetic order predicted by the spin Hamiltonian is in agreement with our RDMFT findings. 
%Let us now discuss some properties of the spin model. For $\gamma = 0$, we expect that the system exhibits N\'eel order for any value of $\alpha$. Both the usual square lattice ($\alpha=0$) and the $\pi$--flux lattice ($\alpha=1/2$) share the Heisenberg antiferromagnet as their effective spin model which shows N\'eel order. For smaller values of $\alpha$ the system is anisotropic and prefers AF order along $z$, which is in agreement with our RDMFT results. In the presence of non-zero $\gamma$  we cannot exclude non--magnetic or even more exotic phases, but the rigorous analysis of the spin model is not our focus here. Within RDMFT we find different types of spiral magnetic order in the $S^y$--$S^z$-plane for $0 < \gamma < 0.25$ and $0 < \alpha < 1/2$ similar to Fig.~\ref{fig:Uc-alpha}(b). Details will be given elsewhere. 

%%%%%%%%%%%%%%%%%%%%%%%%%%%%%%%%%%%%%%%%%%%%%%%%%%%%%%%%%%%%%%%%%%
%%%%%%%%   T I       A T        A L P H A   =   1 / 6   %%%%%%%%%%
%%%%%%%%%%%%%%%%%%%%%%%%%%%%%%%%%%%%%%%%%%%%%%%%%%%%%%%%%%%%%%%%%%

\emph{Topological insulators}.---
Let us now turn to the study of interaction effects on the gapped phases. %In the following we study the effect of interactions on the gapped topological phase. 
For $U=0$, we distinguish the normal (NI) and topological (TI) insulating phases by calculating the $\mathbb{Z}_2$ invariant $\nu$ using Hatsugai's method~\cite{fukui-07prb121403}. %, which is equal to $\nu=0 (1)$ for the NI (TI) phase. 
For $U>0$, we identify the phases by computing the spectral function in a cylindrical geometry using RDMFT and counting the number of gapless helical edge states crossing the bulk gap  (for technical details we refer to the Supplemental Material). The TI phase exhibits an odd number of helical Kramer's pairs per edge while the NI phase an even number (including zero). %,which contains information about edge states in the system. 
Edge states are also crucial for detection of topological phases in cold-atom experiments, and we numerically study how robust they are with respect to interactions. 
%and an external trapping potential. 
In the following, we focus on fixed $\alpha=1/6$, which qualitatively captures all phenomena that occur in this system for general $\alpha = p/q$. 

%\emph{Axial symmetric case}.---
In the axial symmetric case of $\gamma = 0$ there exist TI phases only away from half filling, since the system is a (semi)metal for $n_F = 1$ (and not too large $\lambda_x, U$). This is shown in Fig.~\ref{fig:a16-U0}(a), and is expected as the spinless Hofstadter problem at $\alpha=1/6$ exhibits a QHE with Chern number $C=\pm 2$ for $E_F$ in the two energy gaps closest to zero and a QHE with $C=\pm 1$ for $E_F$ in the other gaps. The Chern number corresponds to the number of chiral edge modes in an open geometry. In the present time--reversal invariant system we thus find an according number of helical Kramer's pairs within the gaps. 

For a filling of $n_F = 1/3, 5/3$ the system is thus a TI. We observe this topological phase to be stable even for large interactions up to $U = 10$. We can induce a NI--TI phase transition in the other gap for $n_F = 2/3, 4/3$ by applying a large enough staggered lattice potential $\lambda_x \geq 1$ [see Fig.~\ref{fig:a16-U0}(a)]. %The staggered potential can deform the bands and for sufficiently strong staggering it creates an additional TI phase at $E_F=\pm 1$; see Fig.\,\ref{fig:a16-U0}\,(b).
%A topological phase appears at $E_F = - 1$ (or $n_F=1/3$)  for $\gamma = 0$ and $\lambda_x \geq 1$ [see Fig.~\ref{fig:a16-U0}(b)]. 
Fixing $n_F = 2/3$ we now turn on interactions, and observe that this phase is quite stable as shown in Fig.~\ref{fig:a16-U0}(b). Eventually, large enough interactions reverse the effect of the staggering potential and drive the system into the NI phase. Note that a static Hartree-like approximation (red dashed line) yields comparable results for small $U$ but overestimates the effect of staggering for larger values of $U$.

%\emph{Broken axial symmetry}.---
A topological phase at half filling occurs only if we break the axial symmetry in the system by considering $\gamma >0$. We present the non-interacting $\lambda_x$--$\gamma$ phase diagram in Fig.~\ref{fig:a16-U0}(c)~\cite{goldman-10prl255302}. The interacting $\lambda_x$--$U$ phase diagram for different values of $\gamma$ is shown in Fig.~\ref{fig:a16-U0}(d). Both semimetal and QSH phases are robust up to interactions of order $U \simeq 3-5$, at which point larger interactions drive the system into a magnetically ordered state. %A qualitative understanding of the interacting phase diagram follows from the observation that interactions mainly reverse the effect of staggering. 
Prominently, we observe an interaction--driven NI to QSH transition for $\gamma = 0.25$ and $\lambda_x \gtrsim 1.5$, and a metal-QSH transition for $0.22 \leq \gamma < 0.25$ and $\lambda_x \gtrsim 1$.    
%In general, we identify the topological phase in the interacting system by computing the spectral function for a cylinder geometry using RDMFT an counting the number of edge states crossing the bulk gap.

Using RDMFT for a cylinder geometry, we are able to directly observe the behavior of the edge states in the interacting system. Gapless edge states are key to different detection schemes of topological phases in cold-atom systems~\cite{top-coldatom, detection}. Since topological phases are uniquely characterized by their helical edge states~\cite{wu-06prl106401},
%~\cite{PhysRevLett.96.106401}, 
a probe of these states is the most direct measurement~\cite{goldman-10prl255302,measurement}. %Relevant to experiments, we have explicitly checked that the edges states are robust with respect to interactions also in the presence of a finite trapping potential $V(\bs{x})\sim |\bs{x}|^\delta$ as long as $\delta \geq 4$~\cite{buchhold-unpub}, which is experimentally feasible.  
In Fig.~\ref{fig:5}, we give an example of the spectral function $A(k_y, \omega)$ for the interaction driven NI-QSH transition at $\gamma = 0.25$, $\lambda_x = 1.5$. For $U = 0.5 $, we find no gapless edge modes that are connecting the two bulk bands, corresponding to NI, while at $U = 2$ we clearly find a single pair of helical edge modes traversing the bulk gap, which corresponds to the QSH phase.

\begin{figure}[t]
  \centering
  \includegraphics[width=.85\linewidth]{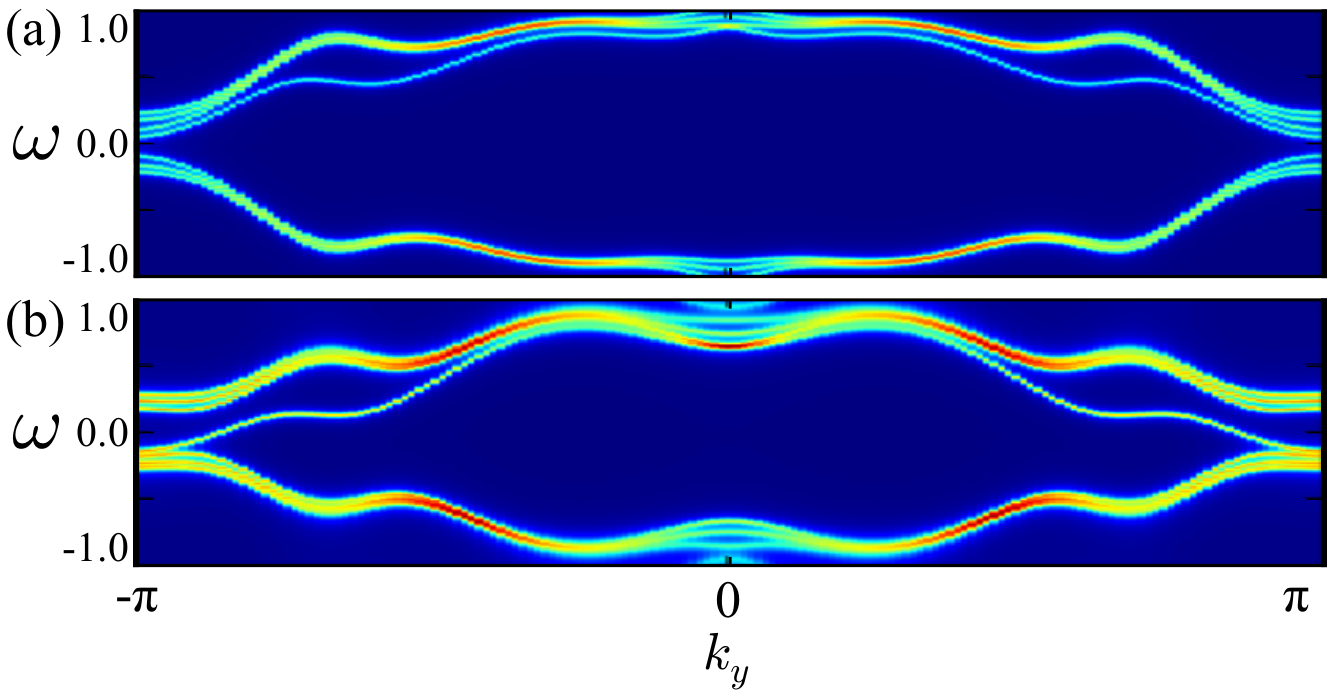}
  \caption{(color online). Spectral function $A(k_y, \omega)$ of interacting system clearly distinguishing between (a) NI phase with no edge states traversing the bulk gap at $U = 0.5$ and (b) QSH phase at $U = 2$ with a single pair of edge modes (per edge) connecting the two bulk bands. Both plots are for $\alpha = 1/6$, $\gamma = 0.25$ and $\lambda_x = 1.5$. }
  \label{fig:5}
\end{figure}

%%%%%%%%%%%%%%%%%%%%%%%%%%%%%%%%%%%%%%%%%%%%%%%%%%%%%%%%%
%%%%%%%%%%%%%%% C O N C L U S I O N %%%%%%%%%%%%%%%%%%%%%
%%%%%%%%%%%%%%%%%%%%%%%%%%%%%%%%%%%%%%%%%%%%%%%%%%%%%%%%%

\emph{Conclusion}.---
We have investigated the TRI Hofstadter-Hubbard model using RDMFT complemented by analytical arguments. 
%use RDMFT to calculate the effect of on-site Hubbard interactions in the spinful and time-reversal-invariant Hofstadter problem with flux $\alpha = p/q$. 
We quantitatively determine the interacting phase diagram including two additional terms available in the cold-atom experiment, a lattice staggering and Rashba-type spin-orbit coupling. Interactions drive various phase transitions. Similar to graphene, we find that a semi-metal at half-filling turns into a magnetic insulator at a critical finite interaction strength. Rashba-type spin-orbit interactions lead to tunable magnetic order with collinear and spiral phases. We explicitly demonstrate the stability of the topological phases with respect to interactions, and verify the existence of robust helical edge states in the strongly correlated TI phase, which is crucial for experimental detection schemes. 

\emph{Note added.}-- After the submission of this work a number of other studies appeared~\cite{spin_studies}  that investigate the strong-coupling limit of fermions (and bosons) in the presence of non-Abelian gauge fields using a spin Hamiltonian similar to our Eq.~\eqref{spin-ham}. 
% which can be exploited for experimental detection of this phase. 

%
%Similar to graphene interactions drive the semi-metal into a magnetically ordered phase at half filling, and the critical interaction strength scales like $U_c \sim q^{-2}$. 
%We show that two additional terms available in the cold-atom experiment that break translational and axial spin symmetry -- a lattice staggering and Rashba-type spin-orbit coupling -- lead to tunable magnetic order with antiferromagnetically and spiral ordered phases. 
%that break translational and axial spin symmetry: a staggering of the optical lattice potential and Rashba-type spin-orbit coupling. 
%We find a rich phase diagram with interactions driving the system into different magnetically ordered phases. 
%We also explicitly demonstrate the stability of the topological phases with respect to weak--to--intermediate interactions, and verify the existence of robust helical edge states in the strongly correlated QSH phase, which can be exploited for experimental detection of this phase. 

\begin{acknowledgments}
The authors acknowledge useful discussions with L. Fritz, K. Sengstock, and I. Spielman. This work was supported by the DFG under Grant No.\ RA~1949/1-1 (S.R.) and via Sonderforschungsbereich SFB-TR/49 and Forschergruppe FOR 801 (D.C., M.B., W.H.), by the NSF under NSF DMR 0803200 (K.L.H.), and the Young Investigator Group of P.P.O. received financial support from the ``Concept for the Future'' of the KIT within the framework of the German Excellence Initiative. K.L.H. and W.H. benefitted from a summer conference at the Aspen Center for Physics supported by the NSF under Grant No. 1066293. D.C. and P.P.O. contributed equally to this work. 

\end{acknowledgments}

\newpage

\title{Supplemental material for ``Time-Reversal-Invariant Hofstadter-Hubbard Model with Ultracold Fermions''}
%%%%%%%%%%%%%%%%%%%%%%%%%%%%%%%%%%%%%%%%%%%%
%
\author{Daniel Cocks}
\affiliation{Institut f\"ur Theoretische Physik, Goethe-Universit\"at, 60438 Frankfurt/Main, Germany}
\author{Peter P. Orth}
\affiliation{Institute for Theory of Condensed Matter, Karlsruhe Institute of Technology (KIT), 76131 Karlsruhe, Germany }
\author{Stephan Rachel}
\affiliation{Department of Physics, Yale University, New Haven, Connecticut 06520, USA}
\affiliation{Institute for Theoretical Physics, Dresden University of Technology, 01062 Dresden, Germany}
\author{Michael Buchhold}
\affiliation{Institut f\"ur Theoretische Physik, Goethe-Universit\"at, 60438 Frankfurt/Main, Germany}
\author{Karyn Le Hur}
\affiliation{Department of Physics, Yale University, New Haven, Connecticut 06520, USA}
\affiliation{Center for Theoretical Physics, Ecole Polytechnique, CNRS, 91128 Palaiseau Cedex, France}
\author{Walter Hofstetter}
\affiliation{Institut f\"ur Theoretische Physik, Goethe-Universit\"at, 60438 Frankfurt/Main, Germany}

\maketitle

\section{Experimental Details}

\subsection{Realizations}

We investigate a model Hamiltonian implementing a non-Abelian artificial
gauge field in a fermionic system. This model system is of general
interest to topological insulator experiments, and directly applicable
to cold-gas systems in optical lattices, of which a wealth of proposals
have been made to implement these artificial gauge fields \cite{Dalibard}. 

Cold-gas experiments can exhibit effective Hamiltonians with artificial gauge 
fields through two main methods: geometric phases or rotation of the cold-gas 
system. A geometry phase is created when a spatially dependent coupling connects 
states of an atom such that the Aharonov-Bohn effect is observed in the 
adiabatic passage of the atom in one particular dressed state through a closed 
loop \cite{Dalibard}. The first such realization of a geometrically induced 
artificial gauge field was performed within a continuous cold-gas of bosonic 
$^{87}\mathrm{Rb}$ with a pair of Raman lasers and a spatially varying magnetic 
field \cite{Lin09}. Alternatively, the Hamiltonian of a rotation system can be 
expressed in the rotating frame, and one finds it is time-independent when the 
system is isotropic and introduces Coriolis terms that take the form of an 
artificial gauge \cite{Fetta}. Such rotating systems have been achieved in the 
laboratory, with features such as a vortex lattice structure \cite{Abo-Shaeer01} 
appearing in bosonic systems.

In a lattice, one can also introduce additional couplings in order to create a 
continuous flux and the so-called ``optical flux lattice'' has been proposed 
\cite{Cooper11} to realize topological non-trivial systems.  However, it is not 
necessary to realize an artificial gauge over the entire lattice. Instead, one 
can make use of the Peierls' substitution \cite{Hofstadter76}, which results in 
a tight-binding model that differs only by a phase factor in either the hopping 
matrix elements or, equivalently, of the local Wannier states. This phase 
factor, the Peierls' phase, can be realized by ``freezing out'' the hopping in a 
lattice and then reconnecting the lattice sites by an additional coupling as 
first proposed by \cite{Jaksch03}. Such a realization of an Abelian artificial 
gauge field for a square
lattice geometry has already been achieved in \cite{Aidelsburger}
by coupling columns of $^{87}\mathrm{Rb}$ in a staggered
optical lattice with Raman couplings, where the artificial magnetic flux 
alternates in sign between neighboring plaquettes. An extension of such a 
lattice to simulate a net magnetic-flux, which exhibits
quantum Hall phases, is proposed in \cite{Gerbier} via the inclusion
of a super-lattice potential. Similarly, \cite{Struck,Hauke} have
shown that it is possible to generate non-Abelian gauge fields by
asymmetric lattice shaking and the extension of any Abelian gauge
field in a cold-gas experiment to a non-Abelian gauge field is possible
by the use of an $n$-state atomic system \cite{Osterloh}. The specific
proposal which we follow, Goldman \emph{et al.} \cite{Goldman_prop},
makes use of this idea by manipulating four states of $^{6}\mathrm{Li}$
to implement a non-Abelian gauge field such that a SU(2) time-reversal
invariant form of the Hofstadter model is realized. With this setup,
it is possible to realize any rational flux ratio $\alpha=p/q<1/2$. For large 
$\alpha$, the edge states associated with these have
been reported \cite{Goldman_prop} to be robust for deviations in
the flux of up to $\Delta\alpha\sim0.01$. In addition, the non-Abelian
nature of the gauge field realizes a spin-orbit coupling, which is
the $\gamma$ term in our Hamiltonian, introduced in equation (3).
This term has a similar effect as the Rashba spin-orbit coupling in
the related Kane-Mele system \cite{Wu}, a model that is closely related to 
graphene. Many other proposals exist
for spin-orbit coupling in continuous ultracold gases \cite{Anderson,Campbell,Cheuk}
which could also be extended to lattice geometries.

\subsection{Detection and Trapping}

The major differences between cold-gas and solid-state experiments
are the mechanisms that are available to observe the topological properties
and edge-states of the system, as well as the size and structure of
the edge that is present. In optical lattices, it is not possible
to perform transport measurements for the Hall coefficients, and so
one must turn to other options to directly observe these states. Several
proposals have been made that use techniques such as Bragg or Raman
spectroscopy \cite{Buchhold,Goldman_edge,Liu}, which directly detect
the edge states and can be applied regardless of the trapping potential,
time-of-flight measurements \cite{Alba} to determine bulk properties,
and direct band mapping by the use of spin-injection spectroscopy
\cite{Cheuk}.

We have investigated detection using Bragg spectroscopy in \cite{Buchhold}
for the non-interacting version of the system considered in this paper.
While it is in principle possible for us to determine the Bragg response via RDMFT
results, we have not observed a qualitative change of the single-particle
spectra corresponding to the edge states. Hence, we do not expect
a qualitative deviation in Bragg response for excitations of these
edge states compared to the non-interacting system.

\section{Herbut's Argument}

In Fig.~1(a) in the main text we presented numerical RDMFT results
for the critical onsite interaction strength $U_{c}(1/q)$ which marks
a zero temperature quantum phase transition between a semi-metallic
phase and a magnetically ordered phase. The system develops a single-particle
gap inside the magnetic phase. The magnetic order is antiferromagnetic
and of N\'eel type. 

In this section, we expand our analytical analysis of the critical
value of $U_{c}$ which we obtain from a slightly modified version
of Herbut's argument. In Ref. \cite{Herbut} I. Herbut considers the
low energy field theory of graphene in the presence of onsite interaction
$U$. Specifically, graphene exhibits four Dirac cones at low energies,
a factor of two stemming from the two valleys ($K,K'$) and a factor
of two stemming from the electronic spin. In his renormalization group
analysis, Herbut extends the electronic spin from two to $N_{s}$
flavors and determines the $\beta$--function of the Hubbard interaction
to leading order in $1/N_{s}$. In his notation, it reads 
\begin{align}
\beta_{a} & =-\tilde{g}_{a}-C_{a}\tilde{g}_{a}^{2}+\mathcal{O}(1/N_{s})\,,
\end{align}
where $\tilde{g}_{a}=g_{a}\frac{2}{N_{s}}$ and $g_{a}=-U/(8\pi\tilde{v}_{F}t)$
with hopping amplitude $t$ and dimensionless Fermi velocity $\tilde{v}_{F}=v_{F}/at$.
Here, $a$ is the short distance cutoff of the theory that is set
such that the Fermi velocity $v_{F}=\tilde{v}_{F}at=1$. The constant
in the $\beta$--function reads $C_{a}=2N_{v}$, where $N_{v}$ is
the number of different valleys. For graphene, one finds that $N_{v}=2$,
but in the spinful Hofstadter system one rather finds that $N_{v}=q$
and so depends on the magnetic flux per unit cell $\alpha=p/q$. The
critical value of the interaction $U_{c}$ is obtained from the condition
that the $\beta$--function changes sign, which yields 
\begin{align}
\left(g_{a}\right)_{c} & =\frac{-2}{C_{a}N_{s}}=\frac{-2}{2qN_{s}}\,.
\end{align}
In terms of the microscopic parameters this reads 
\begin{align}
\left(\frac{U}{t}\right)_{c} & =\frac{16\pi\tilde{v}_{F}}{2qN_{s}}=\frac{4\pi\tilde{v}_{F}}{q}\,,
\end{align}
where we have inserted the physical value of $N_{s}=2$, in our Hofstadter
model. The Fermi velocity $\tilde{v}_{F}(q)$ is exactly known from
the non-interacting band structure of the Hofstadter model, and we
find that it scales as $\tilde{v}_{F}(q)\sim1/q$ for $\alpha=1/q$
in the considered range $2\leq q\leq10$ {[}see inset in Fig.~1(a){]}.
As a result, we predict that 
\begin{align}
\left(\frac{U}{t}\right)_{c} & \sim\frac{1}{q^{2}}
\end{align}
which agrees very well with the numerical RDMFT results as we show
in the inset of Fig.~1(a).

\section{Derivation of effective spin Hamiltonian}
In the main text we introduce the fermionic tight-binding Hamiltonian including interactions and in the presence of the two gauge fields along the $x$ and $y$ links, which reads $H = H_0 + H_I$ with
\begin{align}
    H_0 &= - \sum_j \Bigl\{ t_x c^\dag_{j + \hat{\boldsymbol{x}}} e^{- i 2 \pi \gamma \sigma^x} c_j + t_y c^\dag_{j + \hat{\boldsymbol{y}}} e^{i 2 \pi \alpha x \sigma^z}c_j  + \text{h.c.} \Bigr\} \\
  H_I &= U \sum_{j} n_{j, \uparrow} n_{j, \downarrow} \,.
\end{align}
The summation $j = (x,y)$ with $x,y, \in \mathbb{N}$ runs over all lattice sites of the square lattice with lattice constant set to one. The field operator $c^\dag_{j} = (c^\dag_{j, \uparrow}, c^\dag_{j, \downarrow})$ is a spinor. 

At large interaction, the zeroth order Hamiltonian is given by the interaction part $H_I$. To pursue perturbation theory in the hopping amplitudes $t_{x,y}$ and derive an effective spin Hamiltonian at half-filling up to $\mathcal{O}(t_{x,y}^2/U)$, we follow the analysis in Ref.~\onlinecite{auerbach_quantum_magnetism} and partition the complete Fock space into states with singly occupied sites 
\begin{align}
    S = \bigl\{ \ket{n_{1\uparrow}, n_{1 \downarrow}, n_{2\uparrow}, \ldots} : \forall i : n_{i \uparrow} + n_{i \downarrow} \leq 1 \bigr\}
\end{align}
and states with at least one doubly occupied site
\begin{align}
  D = \bigl\{ \ket{n_{1\uparrow}, n_{1 \downarrow}, n_{2\uparrow}, \ldots} : \exists i: n_{i \uparrow} + n_{i \downarrow} = 2 \bigr\} \,.
\end{align}
We partition the Hamiltonian as 
\begin{align}
  H &= \bmx A & B \\ C & D \emx = \bmx P_S H P_S & P_S H P_D \\ P_D H P_S & P_D H P_D \emx \,,
\end{align}
where $P_S (P_D)$ projects onto the subspace $S (D)$. The effective Hamiltonian in the low-energy subspace $S$ is obtained by projecting the resolvent operator $G(E) = ( E - H)^{-1}$ on this subspace $P_S G(E) P_S = [ E - \mathcal{H}(E)]^{-1}$ with $\mathcal{H}(E) = A + B \frac{1}{E - D} C$. Expanding to lowest order in $t_{x,y}/U$ and $E/U$ yields the (energy independent) low-energy effective Hamiltonian 
\begin{align}
\label{eq:2}
  \mathcal{H} &= P_S H_0 P_S + P_S H_0 P_D \Bigl( - \frac{1}{U} \sum_{j} n_{j \uparrow} n_{j \downarrow} \Bigr) P_D H_0 P_S \,.
\end{align}
At half-filling the first term $P_S H_0 P_S$ vanishes, because starting from subspace $S$ each hopping event of $H_0$ creates a doubly occupied site and the projection onto $S$ yields zero. It requires at least two hopping events for the system to return to the singly-occupied subspace $S$. The second term in Eq.~\eqref{eq:2} contains all possible second-order virtual hopping events. Inserting the hopping Hamiltonian $H_0$ in Eq.~\eqref{eq:2} and defining the $2\times 2$ complex hopping matrices $T_{\hat{\boldsymbol{x}}} = t_x \exp( - i 2 \pi \gamma \sigma^x)$ and $T_{\hat{\boldsymbol{y}}}(x) = t_y \exp(i 2 \pi \alpha x \sigma^z)$, the effective Hamiltonian can be written as
\begin{align}
\label{eq:3}
  \mathcal{H} &= - \frac{1}{U} P_S \Bigl[ \sum_{j,k}  \sum_{\nu, \mu \in \{\pm \hat{\boldsymbol{x}}, \pm \hat{\boldsymbol{y}} \}} (\nu, \mu)_{j,k} \Bigr] P_S \,.
\end{align}
Here, $\hat{\boldsymbol{x}} = (1,0)$, $\hat{\boldsymbol{y}} = (0,1)$ are unit vectors and 
\begin{align}
\label{eq:8}
  (\nu, \mu)_{j,k} &= c^\dag_{j+\nu} T_\nu c_j c^\dag_{k+\mu} T_\mu c_k \,.
\end{align}
Note that the hopping matrices fulfill $T_{-\nu} = T_\nu^*$. To arrive at Eq.~\eqref{eq:3} we have also employed that $P_D \sum_j n_{j,\uparrow} n_{j, \downarrow} P_D = 1$, since exactly one site is doubly occupied in the system after the first hopping event.

To proceed, we use that the system must return to the singly occupied sector $S$ after the second hopping event. Otherwise, the state is annihilated by the projection operator $P_S$. It follows that only terms with $\nu = - \mu$ in the sum in Eq.~\eqref{eq:3} are non-zero. For $\nu = -\mu = \pm \hat{\boldsymbol{x}}$, this demands that $k = j \pm \hat{\boldsymbol{x}}$ in Eq.~\eqref{eq:8}. Explicitly, we find (up to constants)
\begin{align}
  \label{eq:4}
  (\hat{\boldsymbol{x}}, - \hat{\boldsymbol{x}}) & =  t_{x}^2 \sum_{j; \sigma,\sigma' \atop \tau,\tau' } c^\dag_{j+\hat{\boldsymbol{x}}, \sigma} [ \delta_{\sigma \sigma'} \cos (2 \pi \gamma) - i \sigma^{x}_{\sigma \sigma'} \sin ( 2 \pi \gamma) ] \nonumber \\ &  \times c_{j, \sigma'} c^\dag_{j, \tau} [ \delta_{\tau\tau'} \cos ( 2 \pi \gamma) + i \sigma^x_{\tau \tau'} \sin (2 \pi \gamma) ] c_{j+\hat{\boldsymbol{x}}, \tau'} \nonumber \\
&= - 2 t_x^2 \sum_{j} \{ \cos( 4 \pi \gamma) ( S^y_{j+\hat{\boldsymbol{x}}} S^y_j + S^z_{j + \hat{\boldsymbol{x}}} S^z_j ) \nonumber \\ & + \sin(4 \pi \gamma) (S^z_{j+\hat{\boldsymbol{x}}} S^y_j - S^y_{j + \hat{\boldsymbol{x}}} S^z_j ) + S^x_{j+\hat{\boldsymbol{x}}} S^x_{j} \}\,,
\end{align}
where the spin indices $\sigma, \sigma', \tau, \tau' \in \{\uparrow, \downarrow\}$ and we have defined the spin operators $\bfss_j = \frac12 \sum_{\tau \tau'} c^\dag_{j, \tau} \boldsymbol{\sigma}_{\tau\tau'} c_{j,\tau'}$ with $\boldsymbol{\sigma} = (\sigma^x, \sigma^y, \sigma^z)$. We have used the relation 
\begin{align}
  \label{eq:5}
  \sum_{\sigma, \tau} c^\dag_{j,\sigma} c_{k,\sigma} c^\dag_{k,\tau} c_{j,\tau} = - 2 \bfss_j \cdot \bfss_k - \frac12 n_j n_k + n_j \,,
\end{align}
where $n_j = 1$ in our case. Since Eq.~\eqref{eq:4} is purely real, its hermitian conjugate term $(-\hat{\boldsymbol{x}},\hat{\boldsymbol{x}})$ is identical to it. Similarly, we obtain for virtual hopping along the $y$-direction 
\begin{align}
  \label{eq:6}
  (2,-2) &= - 2 t_y^2 \sum_j \{  \cos(4 \pi \alpha x) (S^x_{j + \hat{\boldsymbol{y}}} S^x_j + S^y_{j + \hat{\boldsymbol{y}}} S^y_j ) \nonumber \\ & + \sin(4 \pi \alpha x) ( S^x_{j + \hat{\boldsymbol{y}}} S^y_j - S^y_{j + \hat{\boldsymbol{y}}} S^x_j ) +  S^z_{j + \hat{\boldsymbol{y}}} S^z_j \}
\end{align}
where $j = (x,y)$. The complete low-energy spin Hamiltonian thus reads 
\begin{align}
  \label{eq:7}
  \mathcal{H} &= \frac{1}{U} \sum_{\nu \in \{ \pm \hat{\boldsymbol{x}}, \pm \hat{\boldsymbol{y}}} (\nu, -\nu)  \nonumber \\ 
  &= \frac{t_x^2}{U} \sum_j \biggl\{  S^x_j S^x_{j+\hat{\bfx}} + \cos (4 \pi \gamma) \Bigl[S^y_{j} S^y_{j+\hat{\bfx}} + S^z_{j} S^z_{j+\hat{\bfx}} \Bigr] \nonumber \\ &+ \sin( 4 \pi \gamma ) \Bigl[ S^z_j S^y_{j+\hat{\bfx}} - S^y_{j} S^z_{j+\hat{\bfx}}  \Bigr] \biggr\} \nonumber \\ & + \frac{t_y^2}{U} \sum_{j} \biggl\{ \cos \bigl( 4 \pi \alpha x \bigr) \Bigl[ S^x_{j } S^x_{j+\hat{\bfy}} + S^y_{j} S^y_{j+\hat{\bfy}} \Bigr] + S^z_{j} S^z_{j+\hat{\bf y}} \nonumber \\
&+ \sin( 4 \pi \alpha x) \Bigl[ S^y_j S^x_{j + \hat{\bfy}} - S^x_j S^y_{j + \hat{\bfy}}  \Bigr] \biggr\}  \,,
\end{align}
which is the result given in Eq.~(4) of the main text. 

\section{Numerical Method Details}

\subsection{RDMFT details}

We have used the real-space dynamical mean-field theory (RDMFT) \cite{Snoek}
to investigate numerically the effects of weak, intermediate and strong
interactions. RDMFT operates on a finite realization of a lattice
by the Dyson equation:
\begin{align}
\left[G^{-1}\right]_{ij}=\left[G^{0}\right]_{ij}^{-1}-\Sigma_{ij}
\end{align}
where $i$ and $j$ refer to single sites of the system, $G^{0}$
is the non-interacting Green's function, and we adopt a vector notation,
such that each site is decomposed into its spin-dependent components:
\begin{align}
G_{ij}=\left[\begin{array}{cc}
G_{\uparrow\uparrow,ij} & G_{\uparrow\downarrow,ij}\\
G_{\downarrow\uparrow,ij} & G_{\downarrow\downarrow,ij}
\end{array}\right],
\end{align}
and similarly with the other Green's functions and self-energy. The
core assumption of RDMFT, as with standard DMFT, is that the self-energy
is local $\Sigma_{ij}=\Sigma_{i}\delta_{ij}$.

In the RDMFT procedure, we map each site onto an impurity problem
by integrating out all other degrees of freedom in the lattice. To
solve the impurity problem we use a combination of exact diagonalization
(ED), numerical renormalization group (NRG) \cite{Bulla} and the
continuous-time auxiliary spin solver (CT-AUX) \cite{Gull}. The NRG
solver operates directly at $T=0$ and in real-frequency, which we
use for the cases where $\gamma=0$. The ED and CT-AUX solvers operate
in Matsubara frequencies and at finite temperature. These solvers
are well known in solving the system without the spin-mixing due to
the $\gamma$ term in Hamiltonian (1), and we make some comments about
the inclusion of spin-mixing below.

To directly observe edge states in our system, we must take a finite
system and enforce cylindrical boundary conditions. That is, we set
periodic boundary conditions in the $y$ direction and open (i.e.
fixed) boundary conditions in the $x$ direction. In this case, we
can assume $k_{y}$ is a good quantum number when no symmetry breaking
has occurred. We investigate system sizes up to $48$ sites in the
$x$-direction and have allowed for symmetry breaking with a periodicity
of up to $24$ sites in the $y$-direction. Because we have observed
no other symmetry breaking than AF order in the $y$-direction, we
are able to restrict the number of impurity solver calculations to
that of the lattice size in the $x$-direction. We also investigate
the periodic system to better determine the onset of magnetism without
boundary effects.

\subsection{Solver details}

To extend the standard Anderson impurity model to include spin-orbit
coupling we must include, at minimum, a term that couples the bath
orbitals of opposite spins to the impurity. The complete Hamiltonian
for the impurity is then given by
\begin{align}
  \label{eq:1}
  H_{AIM} &= -\sum_{\sigma}\mu c_{\sigma}^{\dagger}c_{\sigma}+\sum_{l\sigma}\epsilon_{l\sigma}a_{l\sigma}^{\dagger}a_{l\sigma} \\ & +\sum_{l\sigma}\left[V_{l\sigma}a_{l\sigma}^{\dagger}c_{\sigma}+W_{l\sigma}a_{l\bar{\sigma}}^{\dagger}c_{\sigma}+h.c.\right]+Un_{\uparrow}n_{\downarrow} \nonumber  \,,
\end{align}
where $c_{\sigma}$ is an impurity annihilation operator, $a_{l\sigma}$
is a bath annihilation operator, and $\bar{\sigma}$ represents the
opposite spin to $\sigma$. Here, the term involving $W_{l\sigma}$
represents the extension of the impurity model. Due to this term,
the Weiss Green's functions for the impurity model become:
\begin{align}
\mathcal{G}_{\sigma\sigma}^{-1}(i\omega_{n}) & =i\omega_{n}+\mu-\sum_{l}\left[\frac{|V_{l}|^{2}}{i\omega_{n}-\epsilon_{l\sigma}}+\frac{|W_{l}|^{2}}{i\omega_{n}-\epsilon_{l\sigma}}\right]\\
\mathcal{G}_{\sigma\bar{\sigma}}^{-1}(i\omega_{n}) & =-\sum_{l}\left[\frac{V_{l\sigma}^{*}W_{l\bar{\sigma}}+W_{l\sigma}^{*}V_{l\bar{\sigma}}}{i\omega_{n}-\epsilon_{l\sigma}}\right].
\end{align}

Formally, the CT-AUX solver is unchanged by the inclusion of the spin-mixing,
however one must now also track the spin-mixing Weiss Green's functions,
$\mathcal{G}_{\sigma\bar{\sigma}}$, for each configuration of auxiliary
spins. This means that the fast-matrix updates for the solver \cite{Gull}
become rank-2 updates. In addition, we were unable to prove that the
determinants which must be evaluated will always be real, and so we
must consider the possibility of a ``sign problem'' (or more precisely,
a phase problem) in the Monte-Carlo sampling. However, we have carefully
analyzed all of our results and observed purely real and positive
weights for all configurations that we consider.

\subsection{Details about gaps and magnetization}

From our converged RDMFT solutions, we determine several quantities
directly from the output of the impurity solvers, including magnetization,
double occupancy and the density of states at the Fermi edge for each
site of the lattice. To identify the phase, we observe the presence
of a charge gap via several methods: a zero density of states at the
Fermi edge, determined from extrapolation of the Matsubara Green's
functions to $\omega=0$, an exponential decay of the imaginary time
Green's functions, which is fitted to the form $G(\tau)\sim\exp(-\tau\Delta_{\mathrm{sp}})$,
and the analytical continuation \cite{Jarrell} to a real-frequency
spectrum, by first continuing the self-energy \cite{Wang}.

In each case, we extract trends for different system size and temperature
to determine the properties of the infinite system at low temperatures.
However, we do not wish to claim exact results for $T=0$ data, as
this is beyond the capabilities of our solver, and therefore explicitly
show the results for $\beta=20$. We have compared these results to
that of $\beta=50$ for some cases and do not notice significant changes
in the critical interactions for magnetic transitions.

In our system we observe two different phase transitions in the presence
of interactions: a) Semi-metal to magnetic order. In this transition,
we observe a simultaneous opening of the single-particle gap and magnetic
order throughout the entire lattice, and b) Quantum spin-Hall phase
to magnetic order. Here the system always possesses a single-particle
gap in the bulk, and so the only noticeable change in the bulk is
the onset of magnetic order. Near the edge, the topological edge states
also become gapped in this transition, however at larger temperatures
we observe the single-particle gap opening later than the appearance
of magnetization. As the magnetization has destroyed time-reversal
symmetry, these states will be susceptible to disorder and localize.
The fate of these states in ultracold atom experiments is unknown
and requires further investigation.

\end{document}